\documentclass{aa}
\usepackage{graphicx}
\usepackage{txfonts}
\usepackage{xcolor}
\usepackage{placeins}
\usepackage{amssymb}
\usepackage{gensymb}
\usepackage{hyperref}

\begin{document}

\title{Ion-Scale Current Sheets Embedded in Reconnection Jet Shear Layer of the Near-Sun Heliospheric Current Sheet}
\titlerunning{Ion-scale current sheets in the HCS jet shear layer}

\author{Dae-Young Lee\inst{1}
\and Dooyoung Choi\inst{1}
\and Kyung-Eun Choi\inst{2}
\and Sung Jun Noh\inst{3}}

\institute{Department of Astronomy and Space Science, Chungbuk National University, Chungbuk 28644, Korea\\
\email{dylee@chungbuk.ac.kr}
\and
Space Sciences Laboratory, University of California, Berkeley, CA, USA
\and
New Mexico Consortium, Los Alamos, NM, USA; ISR-1: Space Science and Applications, Los Alamos National Laboratory}

\date{}

\abstract
{Magnetic reconnection in the heliospheric current sheet (HCS) plays an important role in restructuring the solar wind magnetic topology and generating plasma jets and magnetic islands. While large-scale signatures of HCS reconnection have been reported in many observational studies, the kinetic-scale structure embedded within reconnection regions remains less well understood.}
{We investigate the ion-scale currents sheets (CSs) embedded within an HCS reconnection region and their relationship to the flow-shear layer at the edge of a reconnection jet.}
{We analyzed an HCS crossing observed by the Parker Solar Probe on March 29, 2024, using high-time-resolution magnetic field measurements. We focused on ion-scale magnetic transitions within two brief intervals of flow-shear layer at the edges of the reconnection jet and examined them in a local LMN coordinate system.}
{Twelve representative CSs are identified, whose duration is on average $\sim$0.06 sec, corresponding to spatial scales of only a few ion inertial lengths. They are classified into three types based on the behavior of the out-of-plane magnetic component $B_M$: (1) CSs showing clear bipolar $B_M$ variations without bifurcation in reconnecting-field ($B_L$), (2) CSs with both bipolar $B_M$ variations and bifurcated $B_L$ profiles characterized by a plateau structure, and (3) CSs where strong fluctuations obscure an otherwise expected bipolar signature.}
{The reconnection jet shear layer in the HCS may serve as an active site that hosts a chain of ion-scale CSs. This provides new insight into the multiscale structure of HCS reconnection and suggests that flow shear layers may play an important role in generating secondary kinetic-scale structures.}


\maketitle

\section{Introduction}

Magnetic reconnection is a fundamental plasma process that converts magnetic energy into bulk flow, plasma heating, and particle energization. In the solar wind, reconnection has been identified in thin current sheets (CSs) over a wide range of heliocentric distances, and direct in situ evidence has been available since the ACE and Wind era \citep[e.g.,][]{Gosling2005,Gosling2006,Lavraud2009,Phan2020,Phan2021,Eriksson2022}.

Among solar-wind CSs, the heliospheric current sheet (HCS) is of particular interest because it is the largest and most persistent polarity-reversal structure in the heliosphere. Its large-scale extent, long-lived nature \citep[e.g.,][]{SanchezDiaz2019,Szabo2020,Choi2024,Choi2025}, and frequent association with plasma-sheet structure make it a natural environment for studying reconnection in an astrophysical collisionless plasma. A few early observational studies identified reconnection exhausts during HCS crossings at 1 au and showed that reconnection can produce disconnected or closed magnetic field topologies in the solar wind \citep{Gosling2005,Gosling2006,Lavraud2009}. More recent Parker Solar Probe (PSP) observations have further established that the HCS in the inner heliosphere can be accompanied by a distinct heliospheric plasma sheet and can host repeated reconnection signatures close to the Sun \citep{Szabo2020,Phan2020,Phan2021,Liewer2024}. Numerical studies likewise indicate that the HCS, viewed as a system of sectored current sheets, is a favorable site for reconnection and can fragment into flux ropes through tearing and related instabilities \citep{Drake2010,Higginson2018,Reville2020,Reville2022}.

At the same time, a separate line of research has shown that the solar wind contains abundant ion-scale CSs. Observations and simulations of turbulence have demonstrated that coherent current layers naturally emerge near the dissipation range, and in situ observations have linked such structures to intermittent heating and kinetic activity. Recent statistical analyses using PSP and Wind have further quantified the thickness and occurrence of proton-scale CSs in both near-Sun and 1 au solar wind \citep{Perri2012,Osman2012,Lotekar2022,Eriksson2022,Franci2015,Vasko2022,Phan2024}.

However, the relationship between HCS reconnection and ion-scale CS formation remains insufficiently explored. Most HCS reconnection studies have emphasized large-scale exhaust signatures, magnetic topology, or flux-rope production, whereas kinetic-scale CS features in HCSs have been addressed only in a few works \citep{Eriksson2022,Phan2024}. It remains to be clarified how ion-scale CSs are organized within the internal structure of an HCS reconnection exhaust, particularly in regions of strong flow shear at the jet boundary.

In this paper, we report PSP observations of multiple ion-scale CSs embedded in the flow-shear layer near the edge of an HCS reconnection jet. Using high-time-resolution magnetic field data, we identify a sequence of sharp magnetic transitions with characteristic durations corresponding to only a few ion inertial lengths. We further classify the events according to their magnetic signatures, including CSs with clear bipolar out-of-plane fields, bifurcated profiles, and cases where rapid fluctuations obscure otherwise expected bipolar structure. These observations indicate that the jet shear layer contains a rich population of secondary kinetic-scale structures rather than a single smooth exhaust boundary. This is the main distinction of the present work from previous HCS reconnection studies focused mainly on large-scale exhaust signatures. The event therefore provides another observational bridge between global HCS reconnection and local ion-scale structuring, with implications for multiscale energy conversion and plasma dissipation in the inner heliosphere.

For the analysis, we use in situ observations from PSP. Specifically, we use DC magnetic field measurements from fluxgate magnetometers (MAG) in the FIELDS instrument suite \citep{Bale2016} and the proton density, velocity, and temperature measurements from the SPAN-ion instrument in the SWEAP instrument suite \citep{Kasper2016,Livi2022}.

The remainder of this paper is organized as follows. In Sect.~\ref{Section2}, we present an overview of the HCS crossing event and its large-scale plasma and magnetic field characteristics. In Sect.~\ref{Section3}, we describe the identification of ion-scale CSs embedded within the jet shear layer of the HCS event and analyze their magnetic structures in detail. Finally, in Sect.~\ref{Section4}, we discuss the implications of these results and summarize the main conclusions.

\section{Overview of reconnecting HCS crossing event}\label{Section2}

\begin{figure}
\centering
\includegraphics[height=1\textheight, width=\columnwidth, keepaspectratio]{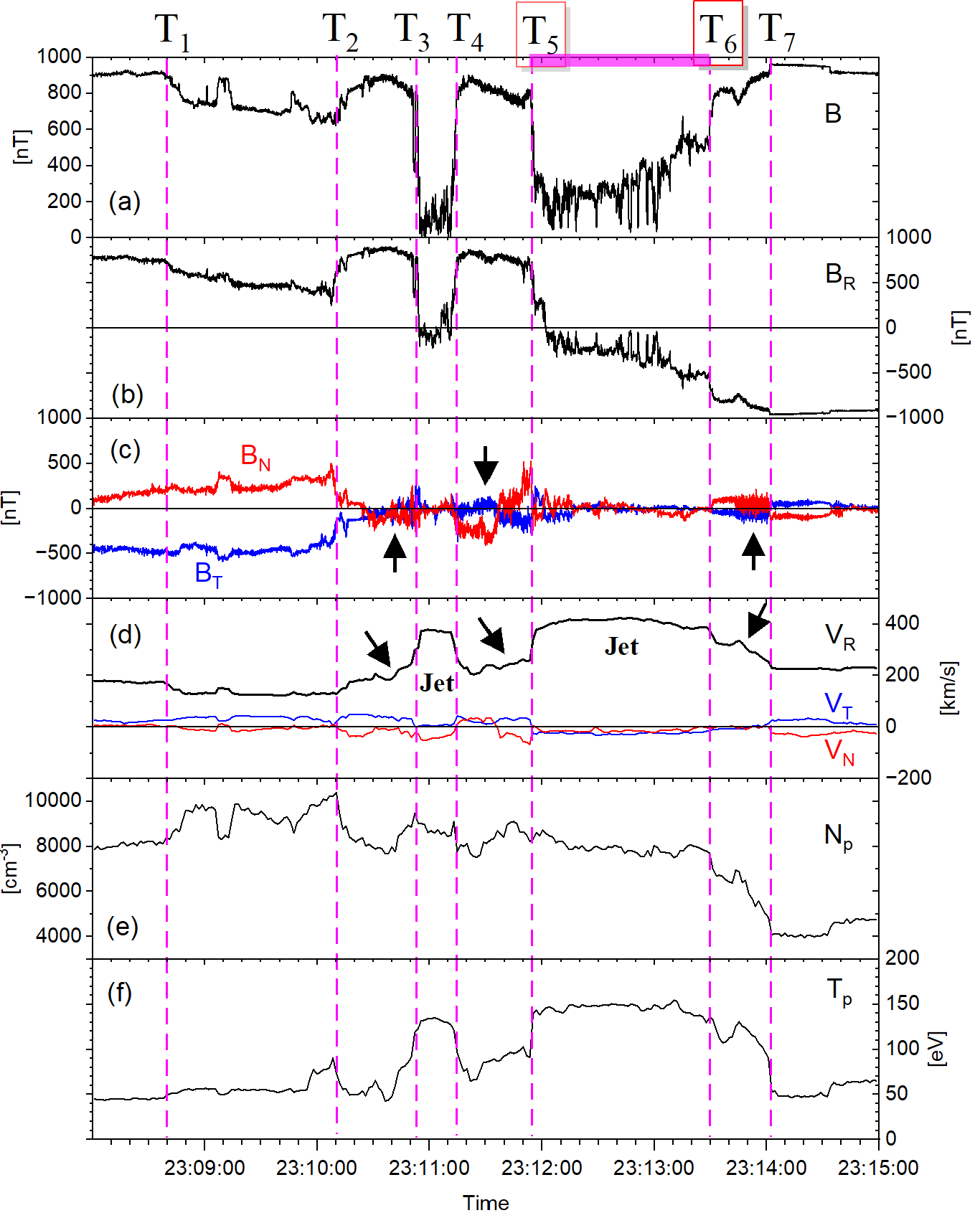}
\caption{Observations of a heliospheric current sheet (HCS) crossing event by Parker Solar Probe (PSP) during an interval on March 29, 2024 at $r \approx 11.6~R_\odot$. (a, b, c) The magnetic field magnitude and magnetic field components in the Radial-Tangential-Normal (RTN) coordinate system, (d) the plasma flow vector components, (e) the proton density, and (f) the proton temperature.}
\label{Fig1}
\end{figure}

Figure~\ref{Fig1} shows an overview of the magnetic field, plasma velocity in the Radial-Tangential-Normal (RTN) coordinate system, and proton density, and proton temperature during the spacecraft passage across the HCS on March 29, 2024. The times $\mathrm{T}_1$--$\mathrm{T}_7$ mark key boundaries identified in the interval.

The spacecraft first enters the HCS at $\mathrm{T}_1$, where the magnetic field magnitude $B$ and the radial component $B_R$ decrease significantly from their upstream values. This transition is accompanied by increases in proton density $N_p$ and temperature $T_p$, consistent with entry into the current-sheet region. The spacecraft finally exits the HCS at $\mathrm{T}_7$, where the magnetic field returns to its asymptotic value of $\sim$900 nT with $B_R$ reversing to the opposite polarity, while both $N_p$ and $T_p$ decrease.

Between $\mathrm{T}_2$ and $\mathrm{T}_3$, the spacecraft briefly retreats toward a region with magnetic field strength close to the asymptotic upstream value. During this interval, the transverse components $B_N$ and $B_T$ exhibit an overall rotation, suggestive of a flux-rope-like structure. A similar retreat toward near-asymptotic field conditions occurs again between $\mathrm{T}_4$ and $\mathrm{T}_5$, where comparable rotations of $B_N$ and $B_T$ are again observed.

Between $\mathrm{T}_3$ and $\mathrm{T}_4$, the spacecraft encounters a partial crossing of the HCS, accompanied by a localized plasma jet (panel (d)) and an increase in proton temperature. The main crossing begins at $\mathrm{T}_5$. We refer to the interval $\mathrm{T}_5$--$\mathrm{T}_6$ (marked by a horizontal bar at the top of panel (a)) as the main HCS crossing. In this interval, the radial magnetic component $B_R$ exhibits an overall bifurcated structure though with some fluctuations (panel (b)), while a strong anti-sunward flow jet develops (panel (d)), together with enhanced proton temperature. These signatures are consistent with a reconnection exhaust embedded within the HCS \citep{Gosling2006,Phan2020,Phan2021}.

Close to the HCS center (particularly during $\mathrm{T}_3$--$\mathrm{T}_4$ and $\mathrm{T}_5$--$\mathrm{T}_6$), large-amplitude fluctuations are observed primarily in $B_R$ and the total field magnitude (panels (a), (b)). In contrast, in the intervals $\mathrm{T}_2$--$\mathrm{T}_3$, $\mathrm{T}_4$--$\mathrm{T}_5$, and $\mathrm{T}_6$--$\mathrm{T}_7$, enhanced fluctuations occur mainly in $B_T$ and $B_N$ (arrows in panel (c)) together with small density decreases (panel (e)) forming density cavities and modest reductions in proton temperature (panel (f)). Such signatures are commonly reported in separatrix regions in both spacecraft observations and kinetic simulations \citep{Shay2001,Mozer2002,Cattell2005,Khotyaintsev2006,Retino2006,Yang2006,Lu2010,Zhou2011,Wang2013}. Finally, in these intervals, the radial velocity $V_R$ shows shoulder-like features (arrows in panel (d)), suggesting that the spacecraft did not reach the fully asymptotic upstream region during the interval.

We determined the orientation of the HCS by estimating the boundary normal using two independent methods. In the first method, the normal vector was obtained from the cross product of the magnetic fields measured just outside the current sheet, $\hat{N}=\mathbf{B}_1\times\mathbf{B}_2/|\mathbf{B}_1\times\mathbf{B}_2|$. Here, $\mathbf{B}_1$ and $\mathbf{B}_2$ represent average magnetic fields in the upstream and downstream asymptotic regions, taken over the intervals 23:00:00--23:05:00 and 23:15:00--23:20:00, respectively. This method yields $\hat{N}=(0.0888, 0.7999, -0.5936)$.

As a second approach, we applied minimum variance analysis (MVA) to the magnetic field data across an interval that includes the entire HCS crossing interval (23:00:00--23:20:00). The resulting normal vector is $\hat{N}=(0.0541, 0.5380, 0.8412)$, with eigenvalue ratios: $\lambda_{\rm max}/\lambda_{\rm int}=34$, $\lambda_{\rm int}/\lambda_{\rm min}=19$, indicating a reasonably well-defined minimum variance direction.

The two normals both lie primarily in the T--N plane in the RTN coordinates. The difference in the sign of the N component is not significant because the MVA normal has an inherent $\pm$ ambiguity. Overall, the angular separation between the two normals is modest, at the level of $\sim$20--$30\degree$, indicating general consistency between the two methods.

We define a hybrid current-sheet coordinate system ($L$, $M$, $N$) by combining the two methods above. Specifically, we determined the local LMN coordinates using the hybrid method as used in \citet{Fargette2021} and \citet{Eriksson2022}. We first adopt the current-sheet normal defined by the cross-product of the magnetic fields on the two sides of the sheet, as explained above. The maximum-variance direction $L_m$ is then obtained from MVA, and the orthogonal axes are constructed as $\hat{M}=(\hat{N}\times L_m)/|\hat{N}\times L_m|$, and $\hat{L}=\hat{M}\times \hat{N}$. This yields a right-handed LMN frame in which $L$ approximates the reconnecting-field direction, $M$ the out-of-plane (guide field) direction, and $N$ the current sheet normal.

Based on this coordinate system, we determined the HCS magnetic shear angle which is $\sim$173$\degree$, indicating a nearly antiparallel configuration. The corresponding guide-field magnitude is $\sim$52 nT. In addition, we evaluated the Wal\'en relation \citep{Paschmann1986} at the exhaust boundaries $\mathrm{T}_5$ and $\mathrm{T}_6$. At $\mathrm{T}_5$, the sense of the plasma velocity jump is opposite to the corresponding Alfv\'en velocity change, with a fitted slope of $\sim$$-0.86$, indicating good Wal\'en agreement. At $\mathrm{T}_6$, the relation is weaker but still broadly Alfv\'enic, with a fitted slope of $\sim$$+0.54$. Thus, the two boundaries show the expected opposite Wal\'en senses for a reconnection exhaust.

\section{Ion-scale current sheets at jet flow shear layers}\label{Section3}

\subsection{Identification and classification of ion-scale current sheets}\label{Section3.1}

In the present work, we focus on two short intervals ($\sim$2 sec each) near $\mathrm{T}_5$ and $\mathrm{T}_6$. We stress that these correspond to the times of the full HCS crossing, when the plasma flow changes significantly and thus mark the jet shear layer. Although the jet flow shear is also present at $\mathrm{T}_3$ and $\mathrm{T}_4$, the interval between $\mathrm{T}_3$ and $\mathrm{T}_4$ correspond to a partial HCS crossing, that is, a short excursion of PSP that did not fully cross the HCS. By focusing on the times of the full HCS crossing, we aim to examine whether ion-scale CSs exhibit asymmetric properties on the two opposite sides of the HCS.

\begin{figure}
\centering
\includegraphics[height=1\textheight, width=\columnwidth, keepaspectratio]{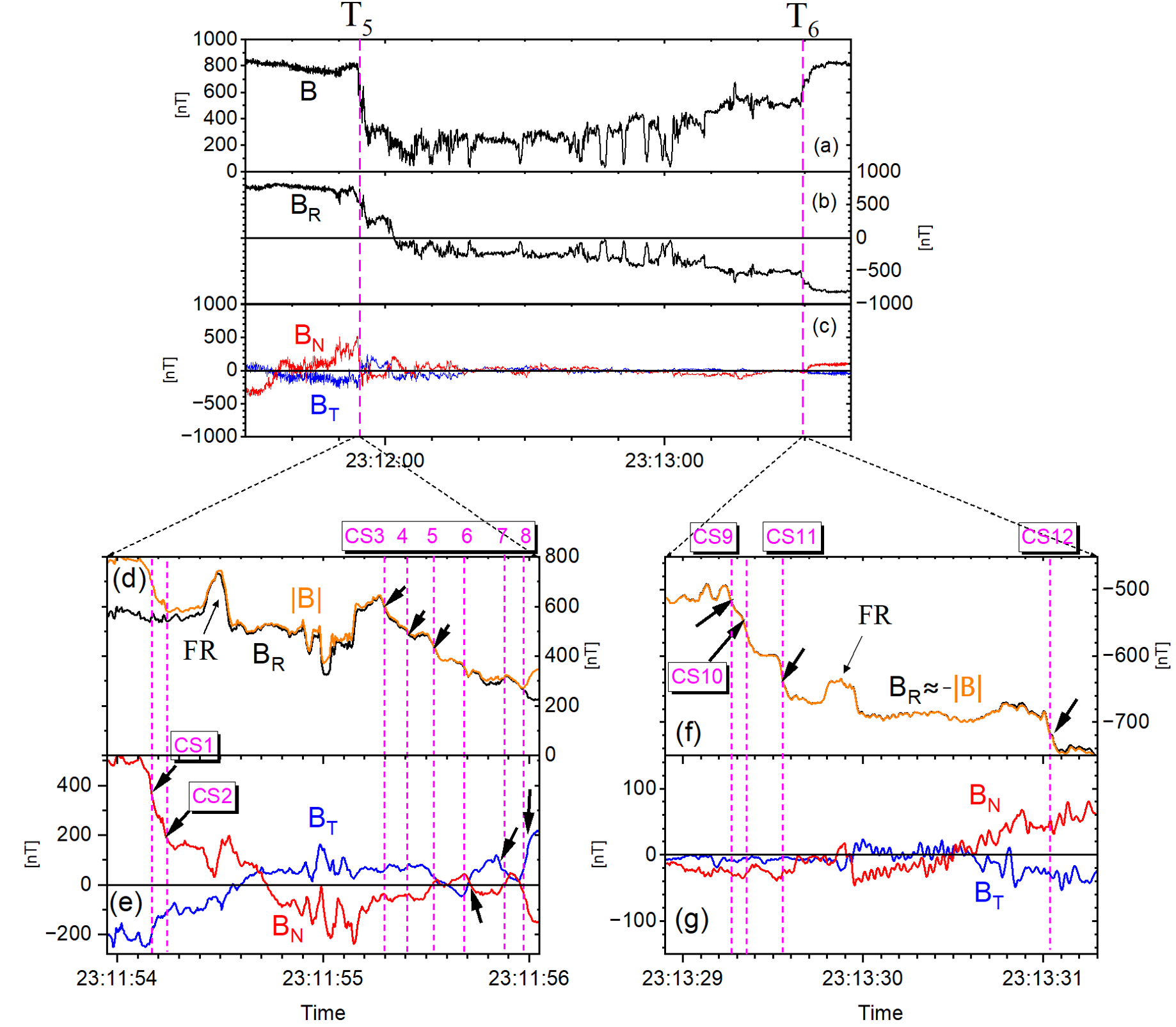}
\caption{Identification of ion-scale current sheets (CSs) at two selected times $\mathrm{T}_5$ and $\mathrm{T}_6$. (a, b, c) The magnetic field magnitude and components for an interval of 2 min and 10 sec surrounding $\mathrm{T}_5$ and $\mathrm{T}_6$. (d, e) Zoom-in view of the magnetic field for $\sim$2 sec interval at $\mathrm{T}_5$. The vertical (magenta) lines mark ion-scale CSs with event identification naming like CS1, CS2, etc. The short arrows mark the main changes in the magnetic field for each CS event. FR in (d) and (f) means flux rope event. (f, g) Zoom-in view of the magnetic field for $\sim$2 sec interval at $\mathrm{T}_6$ in the same format as (d, e).}
\label{Fig2}
\end{figure}

Figure~\ref{Fig2} shows the data for a $\sim$2 min interval surrounding $\mathrm{T}_5$ and $\mathrm{T}_6$. The upper panels (a--c) displays the magnetic field magnitude and RTN components over this $\sim$2 min window, while the lower panels (d--g) present expanded views of the two 2-s intervals.

Within these two intervals, multiple sharp magnetic field variations are evident (panels (d)--(g)). Current sheets are identified visually as abrupt changes in at least one magnetic field component; in particular, we preferentially select a step-like transition observed as PSP crosses the jet shear layers while moving toward and away from the HCS center. Using this criterion, twelve representative CSs (labeled CS1--CS12) are marked by dashed vertical lines in the expanded panels. Detailed magnetic profiles of these CSs are presented in Sect.~\ref{Section3.2} and in Figs.~\ref{Fig3} and \ref{Fig5}--\ref{Fig8}.

The magnetic signatures of the twelve selected CSs vary among events. Changes occur predominantly in $B_R$ for CS3, CS4, CS9, CS10, CS11, and CS12, while CS2 is characterized mainly by variations in $B_N$. In contrast, CS1 and CS5--CS8 involve simultaneous changes in two or three magnetic field components, suggesting more complex local structures.

Based on the local hybrid LMN coordinate system applied to each of the CSs, we estimate CS thickness using $\Delta L=|(V_{\rm sc}-V_{\rm cs})\cdot n|\Delta t$, where $V_{\rm sc}$ is the spacecraft velocity, $V_{\rm cs}$ is the current-sheet velocity, $n$ is the normal of each CS, and $\Delta t$ is each CS transition time. Here we assume $V_{\rm cs}\approx V_{\rm sw}\pm V_A$, where the CS speed relative to the solar wind is limited by the Alfv\'en speed $V_A$. If $V_A$ is neglected, the estimated values of $|(V_{\rm sc}-V_{\rm cs})\cdot n|$ for all CSs range from $\sim$66 to 191 km s$^{-1}$, while $\Delta t$ taken for this estimation lies in the range of $\sim$0.06--0.09 sec. This yields a CS thickness range from $\sim$1.5$~d_i$ to $6.2~d_i$, with an average of $4.5~d_i$, where $d_i\approx2.6$ km is the ion inertial length for $N_p$=8000 cm$^{-3}$. If the CS motion relative to the solar wind is included, using an average $\langle V_A\rangle=128$ km s$^{-1}$ as the upper limit, the estimated thickness range becomes broader, from $\sim$0.6$~d_i$ to $9.4~d_i$, depending on the relative motion of the CS with respect to the solar wind. In contrast, using the spacecraft speed along the HCS normal (obtained from the first method described above in Sect.~\ref{Section2}), $V_{{\rm sc},n}\approx155$ km s$^{-1}$, a 2-s interval near $\mathrm{T}_5$ and $\mathrm{T}_6$, during which twelve CSs were identified, corresponds to a spatial scale of $\sim$310 km, or $\sim$119$~d_i$. Therefore, the identified twelve CSs are ion-scale CSs embedded within an HCS jet shear layer that is roughly two orders of magnitude thicker.

In addition to the twelve CSs which represent the clearest examples of step-like magnetic transitions, additional magnetic field changes exist, such as those around 23:11:55 UT (panels (d) and (e)) and those between $\sim$23:13:30 and 23:13:31 UT (panel (g)), but these are dominated by oscillatory behavior rather than a sharp monotonic change. Incidentally, two short-duration magnetic enhancements are observed and identified as a flux rope (labeled ``FR'' in panels (d) and (f) of Fig.~\ref{Fig2}). These structures exhibit clear rotational signatures primarily in $B_T$ and $B_N$ with durations of $\sim$0.15 sec. The coherent rotation of the transverse field components supports their interpretation as small flux-rope-like structures \citep[e.g.,][]{Moldwin2000,Choi2021} embedded within the jet shear layer.

Based on the behavior of the $B_M$ component in the local LMN coordinate system (described below in Figs.~\ref{Fig3} and \ref{Fig5}--\ref{Fig8}), we classify the twelve CS events into three types.

Type 1 CSs display a clear bipolar variation in $B_M$ (the out-of-plane component) across the sheet, while the main magnetic transition occurs in $B_L$. These events show a relatively simple structure without evidence of bifurcation in the $B_L$ profile. The corresponding CS events are shown in Figs.~\ref{Fig3} and \ref{Fig5}.

Type 2 CSs exhibit a bifurcated structure in the $B_L$ profile as well as a clear bipolar variation in $B_M$. In these events, the $B_L$ transition is characterized by two sharp gradients separated by a plateau region, indicating a double-layer-like or broadened current sheet structure. The corresponding CS events are shown in Fig.~\ref{Fig6}.

Type 3 CSs are characterized by significant fluctuations in $B_M$. These fluctuations obscure any clear bipolar signature that might otherwise be present, making it difficult to identify a well-defined bipolar structure in $B_M$. The corresponding CS events are shown in Figs.~\ref{Fig7} and \ref{Fig8}.

This classification helps organize the variety of ion-scale magnetic structures observed within the HCS interval and provides a framework for examining their detailed properties. We stress again that all three types occur within the jet flow-shear region of the HCS and appear to represent different manifestations of ion-scale CSs embedded in the turbulent reconnection exhaust itself. We also assume that the bipolar perturbations in $B_M$ are possibly Hall magnetic fields.

We emphasize that the average duration of the CSs is only $\sim$0.07 sec and several events exhibit bifurcated profiles or fast transverse oscillations, which provide the basis for our classification into Types 1--3. Because capturing such detailed morphological features is central to the present event-based analysis, we adopted visual selection rather than a purely systematic detection method such as the partial variance of increments (PVI; \citealt[e.g.,][]{Vasko2022}).

\subsection{Magnetic profiles of ion-scale current sheets}\label{Section3.2}

\begin{figure}
\centering
\includegraphics[height=1\textheight, width=\columnwidth, keepaspectratio]{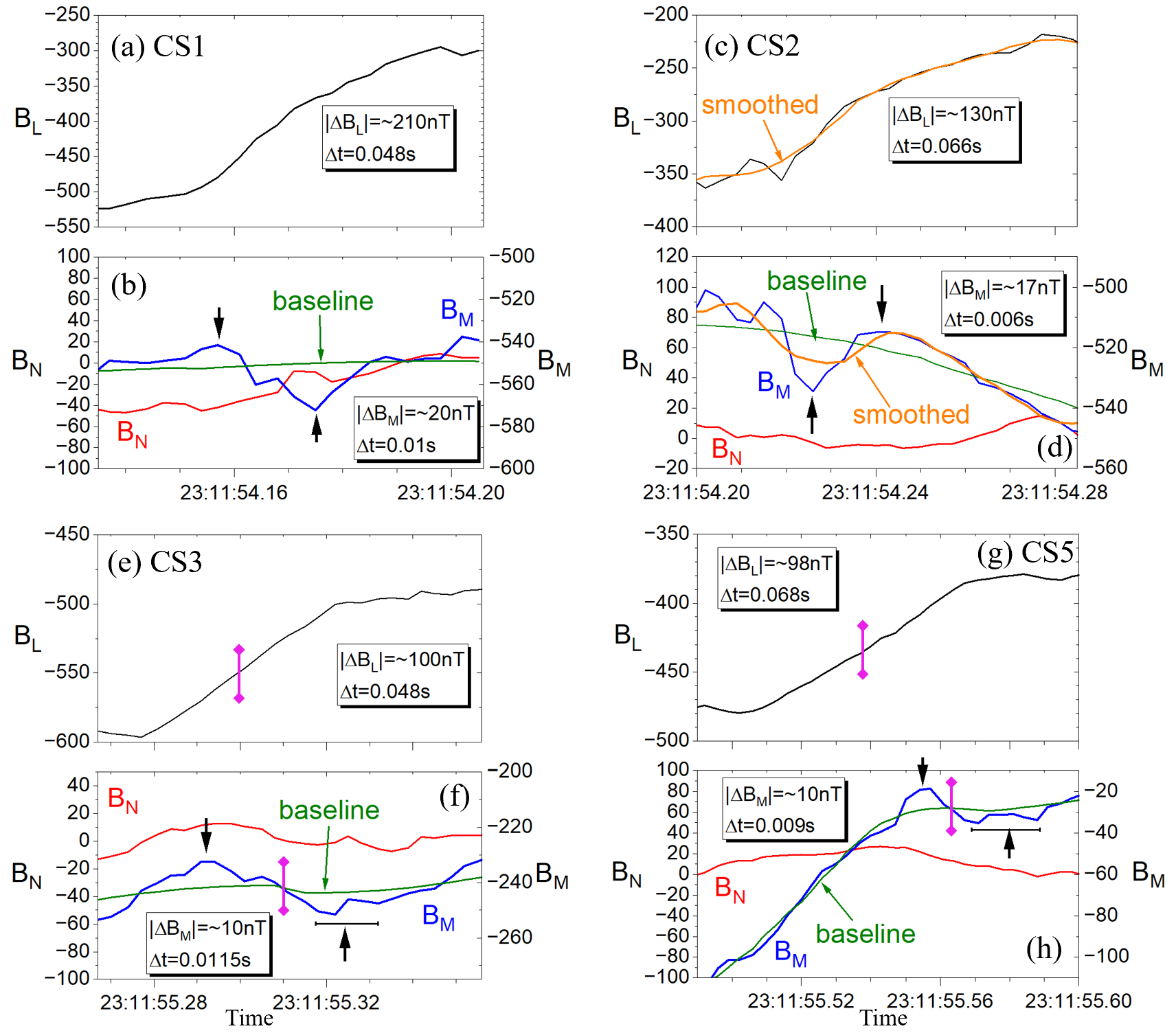}
\caption{A group of four events of ion-scale current sheet crossing as named ``Type 1'': CS1 in (a, b), CS2 in (c, d), CS3 in (e, f), and CS5 in (g, h). The magnetic field components are shown in hybrid LMN coordinates. For each event, baseline (olive) of the M-component, $B_M$ (blue), is provided as a background trend in (b, d, f, h). In (c, d) for CS2, the smoothed curves are further added to help visual identification of variation trends of $B_L$ and $B_M$. The baseline and smoothed curves shown for visual guidance were obtained by applying Savitzky-Golay filter to each current-sheet profile. The short black vertical arrows in panels (b, d, f, h) emphasize the bipolar variation in $B_M$. The short magenta vertical lines in panels (e, f, g, h) mark the approximate center times of the $B_L$ transitions and the associated bipolar variation in $B_M$. The numbers in the insets in the panels for $B_L$ (a, c, e, g) are given for visual guidance about approximate changes in $B_L$ during a major time interval during each CS crossing. Likewise, the insets in the panels for $B_M$ (b, d, f, h) provide approximate rates of the bipolar changes in $B_M$.}
\label{Fig3}
\end{figure}

In this section, we present the magnetic profiles of the twelve CS events, grouped according to the three types defined above. We first consider Fig.~\ref{Fig3}, which presents the magnetic field profiles for four CS events (CS1, CS2, CS3, and CS5) selected from the twelve CSs identified in Fig.~\ref{Fig2}. These CSs belong to Type 1 current sheets, characterized by a well-defined bipolar variation in the $B_M$ component.

For all four events, the dominant magnetic field change occurs in the $B_L$ component, which exhibits a monotonic transition across the current sheet with amplitudes ranging from $\sim$100 to 210 nT (panels (a), (c), (e), (g)). The durations of these transitions range from $\sim$0.048 to 0.068 sec. None of these events show evidence of bifurcation in the $B_L$ profile.

Superposed on these variations, the $B_M$ component (blue in (b), (d), (f), (h)) shows clear bipolar perturbations. The amplitude of each half-cycle of the bipolar signature is on the order of 10--20 nT on a timescale of $\sim$0.006 to 0.0115 sec. In addition, the $B_M$ component exhibits a varying background trend, indicated by the baseline curves (olive in (b), (d), (f), (h)), most prominently in CS2 and CS5. After subtracting this background variation, the bipolar perturbations become more clearly identifiable (short vertical arrows in (b), (d), (f), (h)).

An interesting feature in Fig.~\ref{Fig3} is the offset between the center of the $B_L$ transition and the center of the associated bipolar structure in $B_M$, most notably in CS3 and CS5, as indicated by the short magenta vertical lines in panels (e)--(h). In each case, the peak-to-peak variation in $B_M$ is offset from the midpoint of the $B_L$ transition. This offset suggests that the bipolar perturbation is not perfectly centered on the main field reversal, which may reflect local asymmetries in the current sheet structure or variations in the spacecraft crossing trajectory.

\begin{figure}
\centering
\includegraphics[height=1\textheight, width=\columnwidth, keepaspectratio]{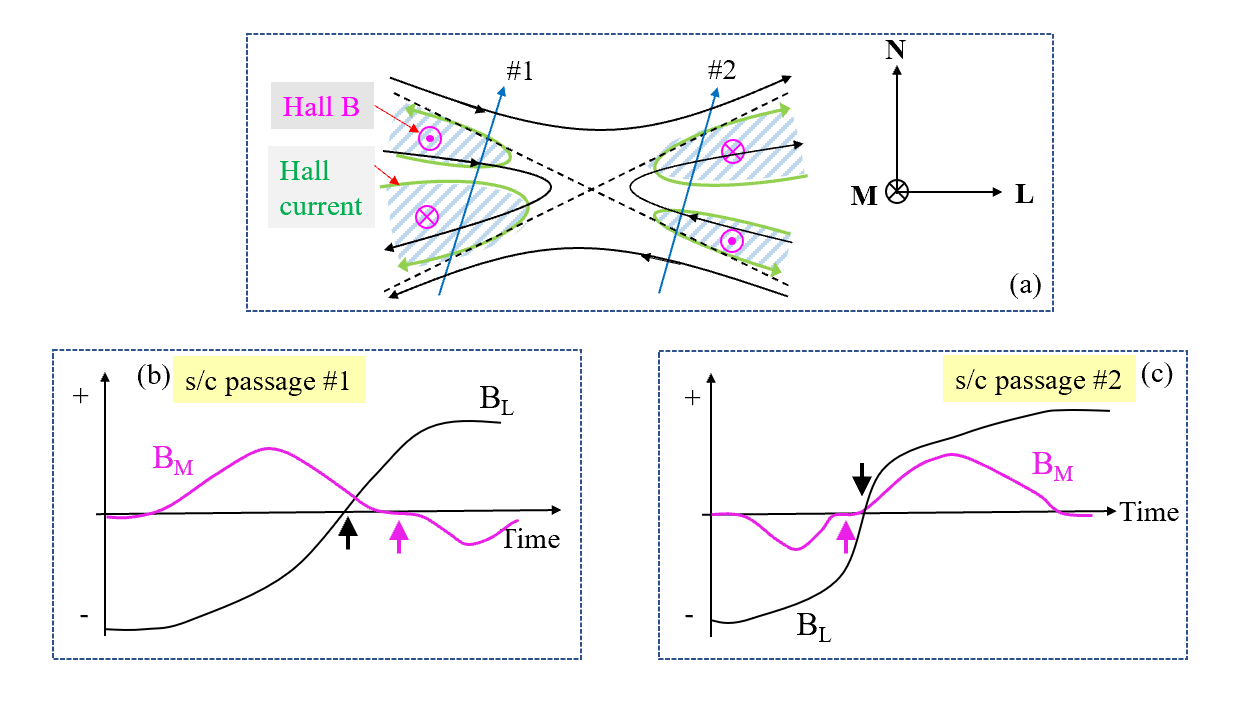}
\caption{Schematic to demonstrate asymmetric nature of bipolar variations in $B_M$ relative to the $B_L$ variation trend in LMN coordinates. The Hall field in (a) is identified as $B_M$ in (b) and (c). The offset between the centers of the variations in $B_L$ and $B_M$ is highlighted by short vertical arrows (black, magenta) in (b) and (c).}
\label{Fig4}
\end{figure}

Figure~\ref{Fig4} presents a schematic illustration of the asymmetric magnetic signatures observed in some of the Type 1 current sheets shown in Fig.~\ref{Fig3}. The schematic is intended only to illustrate the basic geometry and does not attempt to reproduce the detailed structure of individual events. Panel (a) depicts the expected Hall current system and associated Hall magnetic fields in the LMN coordinate system for a reconnecting current sheet. The green arrows indicate Hall currents flowing in opposite directions on the two sides of the current sheet, while the magenta symbols represent the out-of-plane Hall magnetic field ($B_M$). Two example trajectories of spacecraft (\#1 and \#2) crossing the current sheet are indicated by blue arrowed lines. Panels (b) and (c) show the corresponding time series expected for the two example spacecraft crossings. In both cases, while the dominant magnetic transition occurs in $B_L$, $B_M$ exhibits a bipolar variation associated with the Hall magnetic field. Due to the asymmetric Hall current system, the center of the bipolar $B_M$ variation (short magenta arrows in (b) and (c)) is offset from the midpoint of the $B_L$ transition (short black arrows in (b) and (c)).

\begin{figure}
\centering
\includegraphics[height=1\textheight, width=\columnwidth, keepaspectratio]{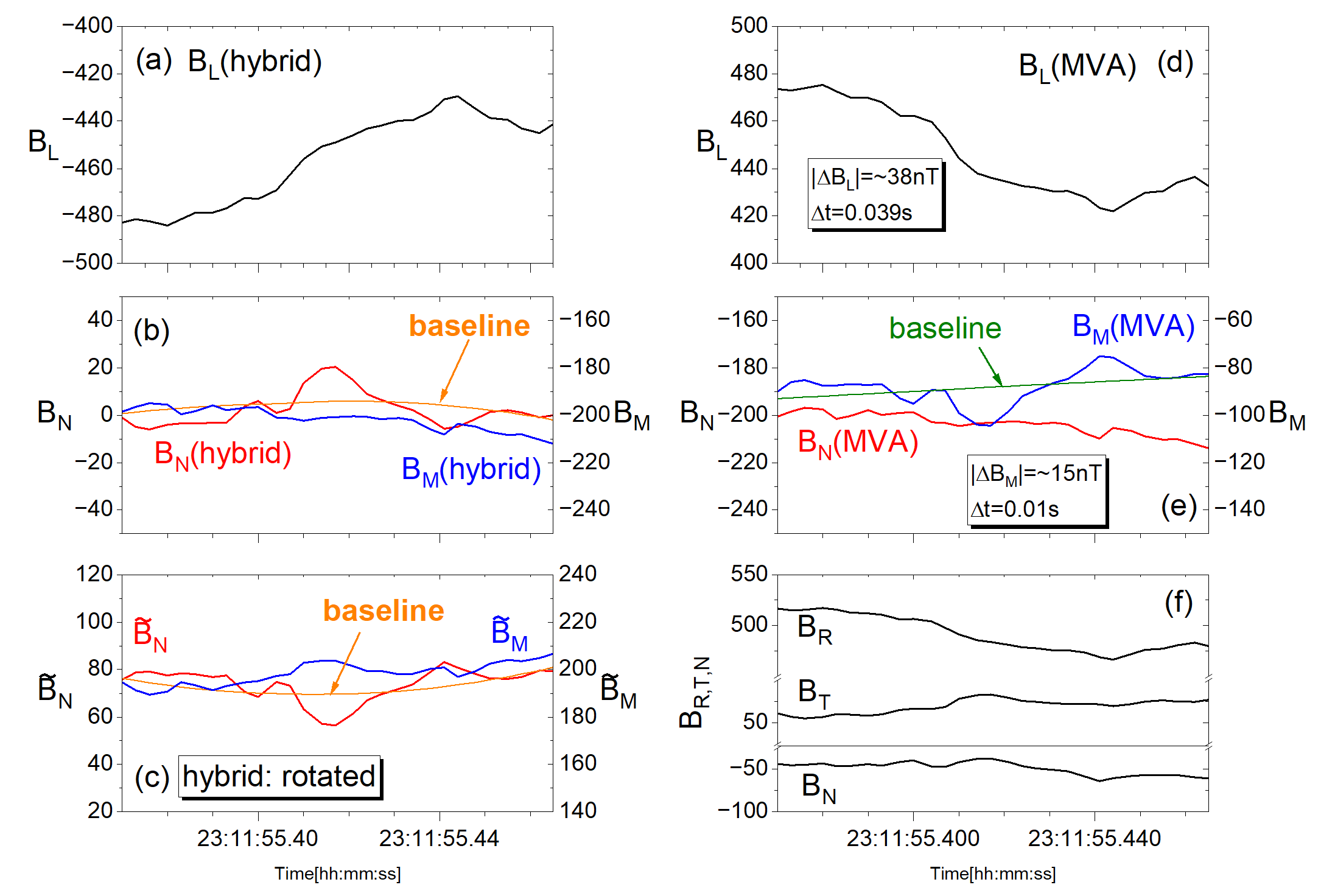}
\caption{Another event of ion-scale current sheet crossing (CS4) belonging to Type 1, comparing the magnetic field variations in different coordinate systems. (a, b) The magnetic field components in hybrid LMN coordinates. (c) Two components obtained by rotation by 159$\degree$ around L-axis in the hybrid LMN coordinate system. (d, e) The magnetic field components in LMN defined by minimum variance analysis (MVA) where L, M, N refer to the maximum, intermediate, and minimum variation directions, respectively. (f) The original magnetic field data in the RTN coordinates.}
\label{Fig5}
\end{figure}

Figure~\ref{Fig5} presents another Type 1 current sheet, CS4. This event is special in the sense that it illustrates how the observed bipolar variation depends on the choice of local coordinates. Specifically, we compare the magnetic profiles between the local hybrid LMN frame and the MVA frame. The event retains the characteristic Type 1 properties in both frames: a well-defined bipolar signature in $B_N$ (hybrid) in panel (b) and $B_M$ (MVA) in panel (e), noticeable asymmetry of these bipolar patterns relative to the center of the main $B_L$ rotation in both coordinate systems, a slowly varying background trend in the transverse component (i.e., the baseline variations in panels (b) and (e)), and no compelling evidence of bifurcation in $B_L$ in either coordinate system (which might exist but cannot be confidently identified). The main difference is that, in the hybrid frame, the bipolar feature appears mainly in $B_N$ rather than $B_M$ in panel (b), whereas in the MVA frame it appears in $B_M$. To clarify this difference, we rotated the transverse components of the hybrid system about the L-axis by $\sim$159$\degree$, choosing the angle so that the rotated transverse direction ($\sim$$B_N$) becomes nearly aligned with the minimum-variance direction $B_M$ (MVA) obtained from MVA. As shown in panel (c), the resulting $\tilde{B}_N$ nearly coincides with $B_M$ (MVA) in the MVA system shown in panel (e). This demonstrates that the rotated system is useful for connecting the hybrid and MVA descriptions and indicates that the different appearance of the bipolar signature in the hybrid and MVA frames arises primarily due to the obliqueness between the hybrid $N$ direction ($-0.03$, 0.73, 0.68) and the MVA minimum-variance direction ($-0.25$, $-0.68$, 0.69).

\begin{figure}
\centering
\includegraphics[height=1\textheight, width=\columnwidth, keepaspectratio]{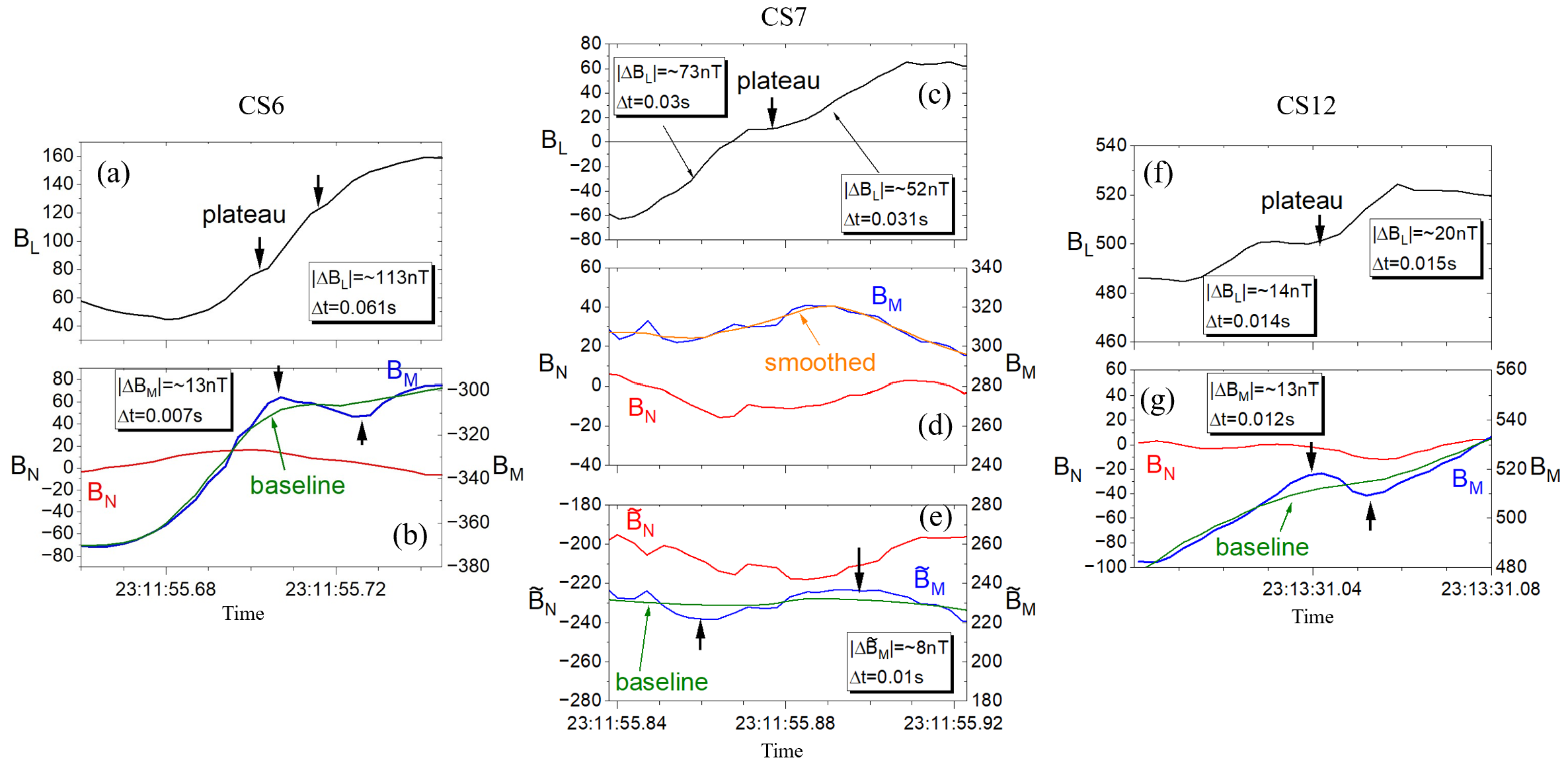}
\caption{``Type 2'' events of ion-scale current sheet crossing that exhibit bifurcation nature with plateau. The magnetic field components are in hybrid LMN coordinates. The magnetic field components in (e) are in the coordinate system rotated by 41$\degree$ around L-axis.}
\label{Fig6}
\end{figure}

Figure~\ref{Fig6} shows the Type 2 current sheets (CS6, CS7, CS12). In contrast to the Type 1 events in Figs.~\ref{Fig3} and \ref{Fig5}, the current sheets in Fig.~\ref{Fig6} are characterized by a plateau in $B_L$. The plateau separates two magnetic-field gradients, suggesting a bifurcated or multi-layered current sheet. CS6 exhibits two plateau intervals (panel (a)), both of which are shorter in duration than the single plateau seen in CS7 and CS12 (panels (c) and (f)). For CS7, the bipolar variation is more clearly revealed in $\tilde{B}_M$ in panel (e), which is obtained after rotating the hybrid LMN transverse coordinates by $\sim$41$\degree$ around the L-axis. This indicates that the Hall (guide-field-direction) component lies between the $M$ and $N$ directions of the hybrid LMN coordinate system. The baseline trends in the varying background $B_M$ or $\tilde{B}_M$ shown in panels (b), (e), and (g) are included to emphasize the localized bipolar perturbations.

\begin{figure}
\centering
\includegraphics[height=1\textheight, width=\columnwidth, keepaspectratio]{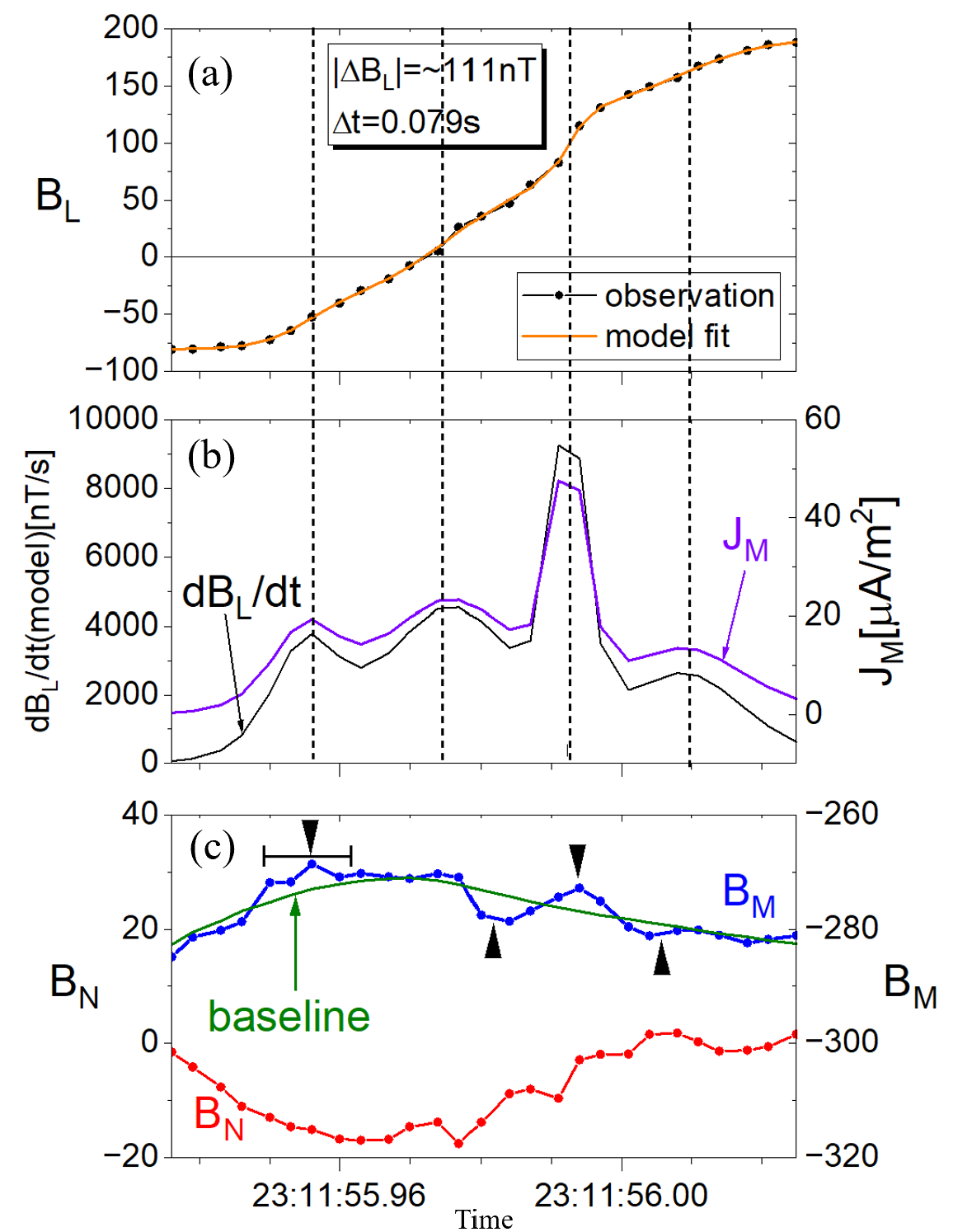}
\caption{``Type 3'' event of ion-scale current sheet crossing (CS8), exhibiting fast-variations in $B_M$ (symbols in (c)), which mask possible existence of the bipolar nature in $B_M$ that would otherwise reveal clearly. In (a) model fitted $B_L$ is shown for reference, which was obtained based on a multiple-Harris sheet model. The time variation rate in $B_L$ and corresponding current density estimation with $V_N=155$ km s$^{-1}$ are shown in (b). All data are shown in hybrid LMN coordinates.}
\label{Fig7}
\end{figure}

Figure~\ref{Fig7} shows an example of Type 3 CSs, CS8. Although the overall magnetic-field rotation in $B_L$ can still be reasonably described by a smooth model profile (yellow in panel (a)), the transverse components exhibit multiple localized perturbations (panel (c)). The inferred current density $J_M$ derived from $dB_L/dt$ (panel (b)) indicates four distinct peaks, suggesting that the structure consists of multiple closely spaced current layers rather than a single monolithic sheet. Correspondingly, in panel (c), $B_M$ (blue) shows several localized variations (triangle symbols) relative to its baseline trend (olive). These multiple perturbations likely mask the bipolar signature of the Hall field that would otherwise be expected for an isolated current sheet.

\begin{figure}
\centering
\includegraphics[height=1\textheight, width=\columnwidth, keepaspectratio]{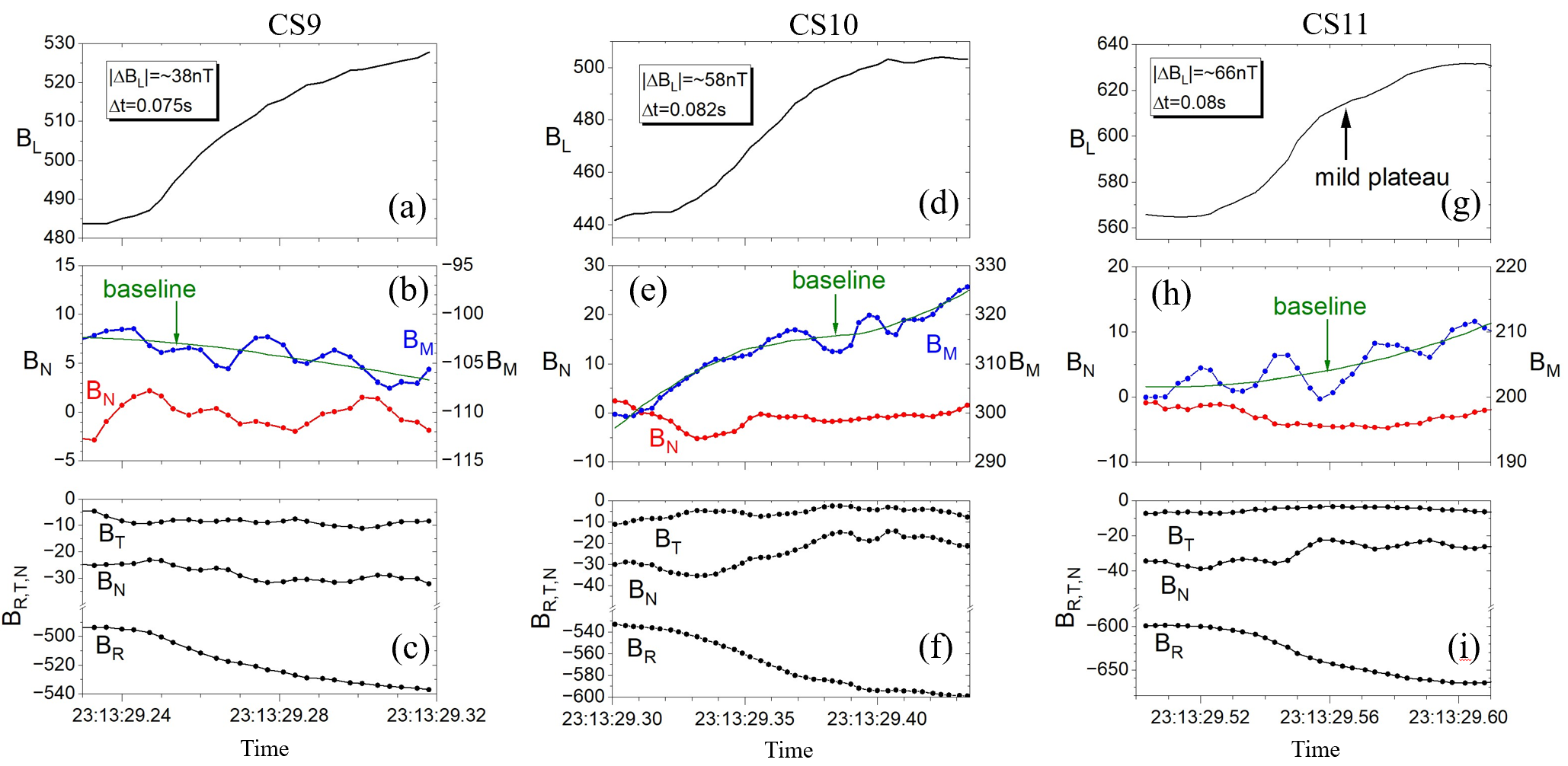}
\caption{Three ``Type 3'' events of ion-scale current sheet crossing (identified at $\mathrm{T}_6$), exhibiting fast variations in $B_M$, which mask possible existence of the bipolar nature in $B_M$ that would otherwise reveal clearly. The bottom panels (c, f, i) show the original magnetic field data in the RTN coordinate system while the data in all the other panels are in hybrid LMN coordinates.}
\label{Fig8}
\end{figure}

Figure~\ref{Fig8} shows the remaining Type 3 current-sheet events (CS9--CS11), which exhibit even stronger oscillatory variations in the transverse magnetic components, $B_M$ and $B_N$, shown in panels (b), (e), and (h). In these cases, $B_M$ displays pronounced oscillations rather than a simple bipolar perturbation. Examination of the magnetic-field components in RTN coordinates shown in panels (c), (f), and (i) indicates that similar oscillations are present primarily in $B_N$ and more weakly in $B_T$, while the magnetic field magnitude is dominated by $B_R$. This implies that the fluctuations occur mainly perpendicular to the background magnetic field, consistent with transverse magnetic oscillations. The oscillations make it difficult to determine whether a true Hall-field bipolar signature is present.

The characteristic oscillation period in the transverse magnetic components, $B_M$ and $B_N$, shown in panels (b), (e), and (h) in Fig.~\ref{Fig8} is approximately 0.02--0.03 sec ($\approx$33--50 Hz). For the local magnetic-field strength ($B\sim$500--650 nT), the proton gyrofrequency is 7.6--9.9 Hz, corresponding to gyroperiods of 0.10--0.13 sec. The observed fluctuations therefore occur well above the MHD regime and may represent Doppler-shifted ion-cyclotron-range waves or related kinetic-scale fluctuations, which will require a more detailed analysis in a separate study.

These results in Type 3 events suggest that the expected Hall-field bipolar signature can be masked by faster transverse magnetic fluctuations in a non-uniform guide field. Alternatively, the Hall structure itself may be multi-layered, so that multiple adjacent Hall-current layers appear as oscillatory variations when the spacecraft traverses the structure.

Taken together, the three types of events suggest that ion-scale CSs in the reconnection outflow region exhibit a range of internal structures rather than a single canonical form. Type 1 events appear closest to the expected picture of a single current sheet with a relatively simple Hall-field perturbation, producing a clear bipolar signature in the transverse magnetic component. Type 2 events show evidence of bifurcated or multi-layered current sheets, indicated by plateau structures in $B_L$ as well as a clear bipolar signature in the transverse magnetic component. Type 3 events exhibit more complicated behavior, where strong transverse oscillations occur at frequencies well above the MHD range and obscure any simple bipolar structure. These results suggest that the Hall-field signature may not always appear as a single bipolar perturbation; instead, the current layer can become structured, fragmented, or embedded in kinetic-scale fluctuations, likely reflecting the dynamic and turbulent nature of the reconnection jet layer.

Additionally it is noteworthy that within the $\mathrm{T}_5$ (entry) interval, the sequence of events (CS1--CS8) tends to evolve from Type 1 to Type 3, suggesting an increase in structural complexity as PSP moves closer to the main body of the jet. By contrast, the $\mathrm{T}_6$ (exit) interval is dominated by the three Type 3 CSs at its beginning, followed by a Type 2 CS event near its end. This pattern indicates a clear asymmetry in the distribution of CS types across the HCS.

\subsection{Current density and orientation of ion-scale current sheets}\label{Section3.3}

\begin{figure}
\centering
\includegraphics[height=1\textheight, width=\columnwidth, keepaspectratio]{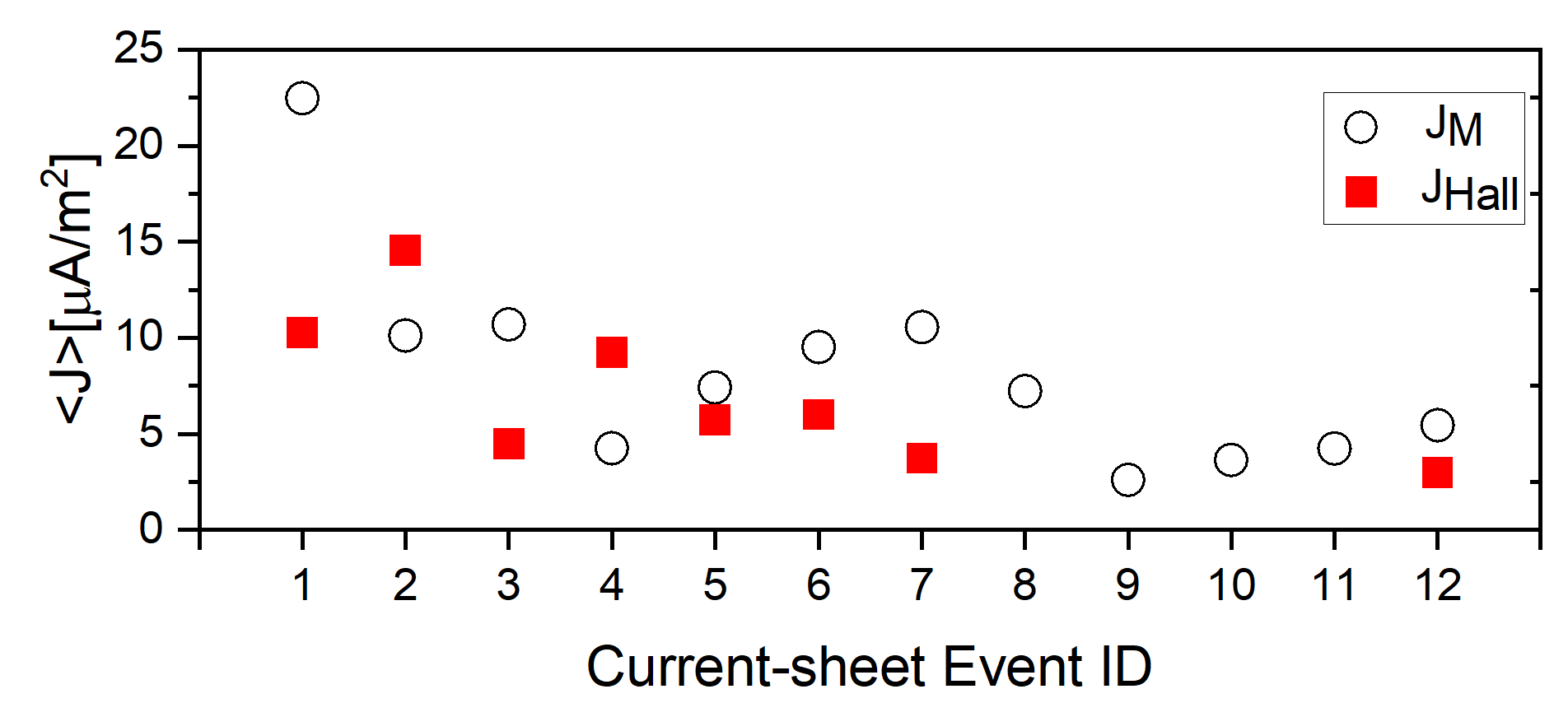}
\caption{Comparison of current densities estimated for all 12 ion-scale current sheet events. $J_M$ and $J_{\rm Hall}$ refer to the current densities corresponding to the changes in $B_L$ and the bipolar changes in $B_M$, respectively.}
\label{Fig9}
\end{figure}

We estimated the average current densities for the 12 CS events and Fig.~\ref{Fig9} summarizes the results. The black symbols ($J_M$) in Fig.~\ref{Fig9} represent the current densities inferred from the main magnetic-field rotation in $B_L$, while the red symbols ($J_{\rm Hall}$) correspond to the current densities associated with the bipolar perturbations in $B_M$, interpreted as the Hall-field signature. Hall-current estimates were not done for CS8--CS11 due to strong oscillatory fluctuations in $B_M$ that obscure a clear bipolar signature. The estimates are obtained using the observed field changes and the spacecraft traversal time across each CS. The comparison shows that the Hall current $J_{\rm Hall}$ is often comparable to, and in some cases a substantial fraction of, the current associated with the guide-field direction $J_M$. This indicates that Hall-scale currents contribute significantly to the internal current structure of these ion-scale CSs, suggesting that kinetic-scale processes play an important role in shaping their structures.

\begin{figure}
\centering
\includegraphics[height=1\textheight, width=\columnwidth, keepaspectratio]{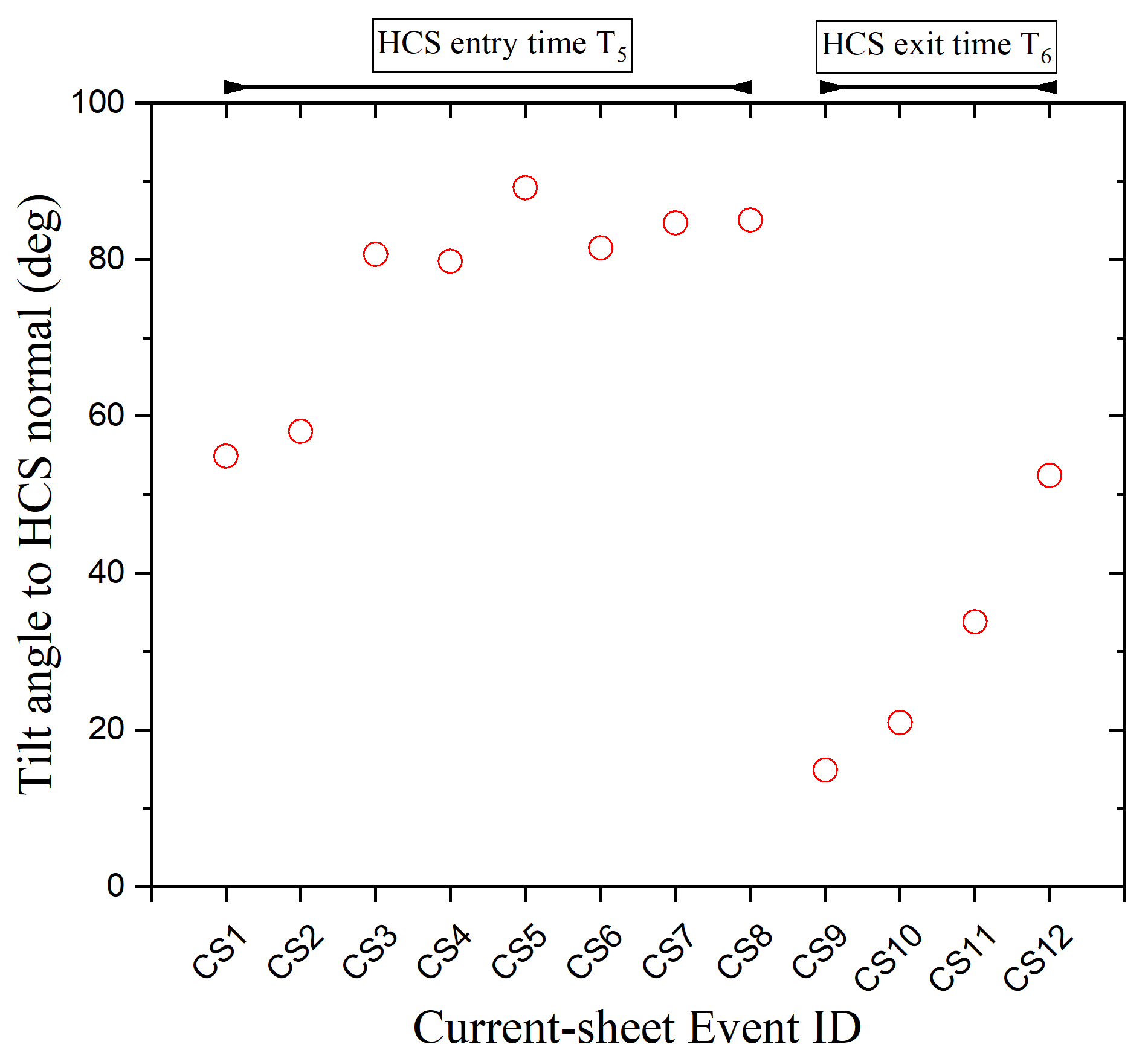}
\caption{Tilt angles of the normal vectors of all 12 ion-scale current sheets (CSs) relative to the HCS normal.}
\label{Fig10}
\end{figure}

We further estimated the tilt angles of the CS normals relative to the HCS normal, and the results are summarized in Fig.~\ref{Fig10}. A clear trend is identified that the CSs observed at the entry time $\mathrm{T}_5$ exhibit systematically larger tilt angles than those observed at the exit time $\mathrm{T}_6$. This difference suggests that the CSs at $\mathrm{T}_5$ are more strongly distorted relative to the large-scale HCS geometry, while those at $\mathrm{T}_6$ likely reflect a relatively more stable configuration closer to the background HCS orientation. Such distortions may be related to the different levels of jet flow shear at the two times, as discussed further in Sect.~\ref{Section3.4}.

Lastly, we evaluated the magnetic shear angle for each CS event. Most of the identified events turned out to be low-shear structures with a substantial guide field. However, these ion-scale CSs are embedded within the strong large-scale magnetic field of the surrounding HCS. In such a situation, the measured field across a given ion-scale CS may reflect a superposition of the local CS-associated field variation and the background HCS field. If the background field remains substantially larger than the reversing component intrinsic to the ion-scale CS, the total magnetic-field vectors on the two sides can differ only weakly in direction, even when the local ion-scale CS itself contains a stronger reversal. In that sense, the strong ambient HCS field may partially mask the intrinsic shear of the embedded CSs. To assess this possibility quantitatively, we carried out an exploratory local background subtraction using the full-cadence magnetic-field data around $\mathrm{T}_5$ and $\mathrm{T}_6$. After removing a smooth local background trend and recomputing the shear angles in the hybrid LMN frame, the inferred shear often increased substantially relative to the raw values obtained from the original full field. In several events, the corrected shear increased from only a few degrees to several tens of degrees, and in some cases to more than 100$\degree$.

Therefore, for at least some events, the strong large-scale HCS field can partially mask the intrinsic field reversal of the embedded CSs. However, for some other events, the estimated shear angles showed appreciable sensitivity to the adopted background model (that is, how the background is defined, including the choice of fitting window and whether a linear or quadratic trend was used). Thus, this result should be presented as a suggestive diagnostic, not yet a unique determination. Nevertheless, it suggests that the low raw shear angles likely reflect a combination of two effects: masking by the surrounding HCS-edge field and the presence of intrinsically low-shear or guide-field-dominated local CSs.

\subsection{Association with jet flow shear}\label{Section3.4}

\begin{figure}
\centering
\includegraphics[height=1\textheight, width=\columnwidth, keepaspectratio]{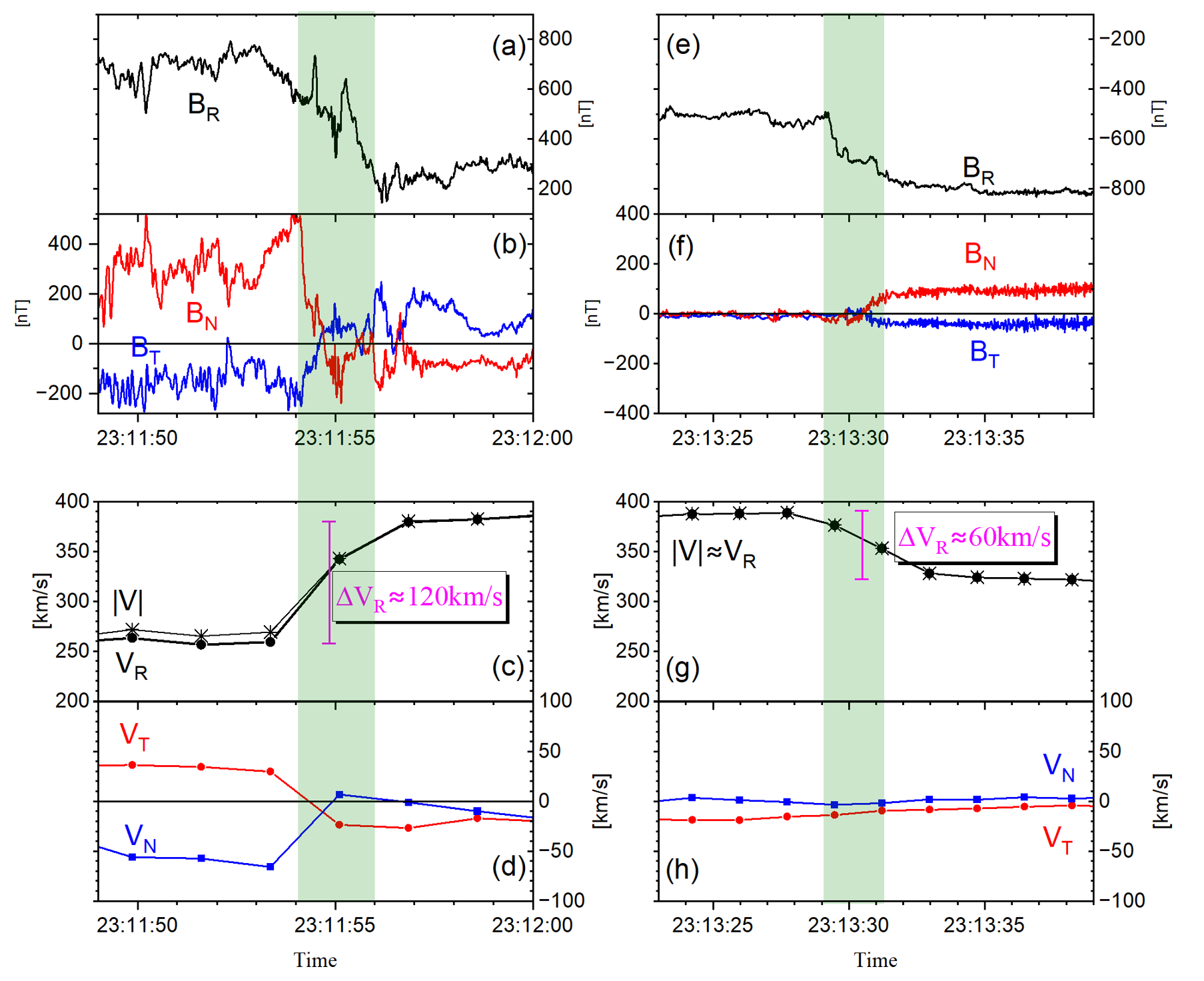}
\caption{Association of HCS crossings with plasma flow shear at (left) HCS entry time ($\mathrm{T}_5$) and (right) HCS exit time ($\mathrm{T}_6$). All data are in the RTN coordinates. The highlights approximately mark the main flow shear layer where the 12 ion-scale current sheets were identified in Fig.~\ref{Fig2}.}
\label{Fig11}
\end{figure}

Here we emphasize that all twelve ion-scale CSs are located within the jet flow shear layer. Figure~\ref{Fig11} shows the magnetic-field and plasma velocity variations across the two intervals. At the entry time $\mathrm{T}_5$ (left column), a substantial change in the radial flow velocity is observed, with $\Delta V_R\approx120$ km s$^{-1}$ (panel (c)). Using the local Alfv\'en speeds estimated from the magnetic-field strengths on each side of the layer, the normalized shear becomes $\Delta V_R/V_A\approx0.8$ on the high-$B$ side and $\Delta V_R/V_A\approx1.6$ on the low-$B$ side. In contrast, at the exit time $\mathrm{T}_6$ (right column), the velocity change is smaller, $\Delta V_R\approx60$ km s$^{-1}$, corresponding to $\Delta V_R/V_A\approx0.5$ on the low-$B$ side and $\Delta V_R/V_A\approx0.4$ on the high-$B$ side.

Since the Kelvin-Helmholtz (KH) instability typically requires $\Delta V/V_A \gtrsim 2$ for efficient growth in strongly magnetized plasmas, the conditions at $\mathrm{T}_5$ approach the marginal regime for KH development, whereas the shear at $\mathrm{T}_6$ is substantially smaller and unlikely to drive KH activity. This difference is consistent with the larger tilts of the CS orientations observed at $\mathrm{T}_5$ compared with the more stable configuration seen at $\mathrm{T}_6$ in Fig.~\ref{Fig10} above.

\section{Discussion and Conclusion}\label{Section4}

The key new element of this study is not merely the detection of ion-scale CSs, but their localization within the flow-shear boundary of an HCS reconnection jet. This event provides an observational example of multiscale structuring, linking an HCS-scale reconnection exhaust to embedded ion-scale CSs and flux-rope-like features in a thin layer. Our results suggest that reconnection-driven jet shear layers in the HCS may act as formation sites for secondary kinetic-scale CSs. A statistical investigation will be needed in future work to evaluate how general and significant this interpretation is.

Other than the KH instability, a possible mechanism for the formation of the observed ion-scale CSs is shear-driven turbulence associated with the reconnection jet boundary. In the present event, the velocity shear layer at the edge of the reconnection jet has a characteristic scale of $L_{\rm shear}$ $\sim$100$~d_i$, which lies in the MHD regime, whereas the thickness of the identified CSs is only a few ion inertial lengths ($L_{\rm CS}$ $\sim$ a few $d_i$). This large scale separation suggests that the MHD-scale shear layer may host a cascade of smaller structures that evolve toward ion kinetic scales. Velocity shear, even in cases where the KH instability is not fully developed, can stretch and fold magnetic field lines locally, thereby amplifying magnetic gradients primarily in directions perpendicular to the mean field and favoring the formation of sheet-like magnetic structures. As nonlinear interactions develop, energy injected at the large shear-scale can cascade through intermediate-scale eddies and generate intermittent current sheets. Such a cascade is expected to terminate near the ion inertial scale, where ions begin to decouple from the magnetic field and Hall effects become important.

In addition, the orientation of the ion-scale CSs appears to be related to the local strength of the flow shear. The HCS normal provides a preferred large-scale direction; however, in the reconnecting HCS studied here, the normals of ion-scale CSs associated with stronger shear near the jet-entry region are systematically rotated away from the projected HCS-normal direction, whereas those observed under weaker shear near the jet-exit region remain closer to it. This behavior is consistent with the expectation that shear-driven turbulence can distort the local magnetic geometry and influence the orientation of small-scale CSs embedded within the reconnection jet shear layer.

On the other hand, it has long been recognized that secondary reconnection within large-scale current sheets can be driven by processes such as outflow turbulence or plasmoid instability, leading to the formation of multiple small-scale magnetic structures within reconnection regions \citep{Matthaeus1986,Drake2006,Servidio2009,Lapenta2015,Comisso2016,Dong2018,Arro2020,Arnold2021,Eriksson2022,Phan2024,Desai2025,Lee2026}. In particular, particle simulations by \citet{Drake2006} showed that when secondary reconnection develops in the presence of a guide field, secondary islands can grow to finite sizes before merging with larger magnetic islands. Similarly, recent simulations by \citet{Desai2025} suggest that the HCS can host multiple reconnection sites, producing numerous small flux ropes that dynamically merge as the system evolves. In such scenarios, reconnection proceeds through the formation of many small magnetic structures that subsequently coalesce into larger flux ropes.

Our conjecture is that these two processes are not necessarily independent. Shear-driven turbulence generated in the reconnection jet boundary can naturally produce intermittent thin CSs through stretching and folding of magnetic field lines. Once such sheets reach ion kinetic scales, they may themselves become unstable and undergo secondary reconnection. In this sense, shear-driven turbulence and secondary reconnection can operate together within the reconnection jet shear layer, with turbulence generating ion-scale CSs and reconnection acting within them to produce Hall fields and flux-rope-like structures. The ion-scale CSs with bipolar signatures and flux-rope-like signatures observed in the present HCS event may therefore represent different manifestations of a multiscale reconnection environment developing within the HCS jet shear layer. Nevertheless, direct observational confirmation of flow jets associated with these ion-scale CSs is not possible because the cadence of the PSP plasma data is insufficient to resolve such short-duration structures.

In the present work, we have focused on ion-scale CSs at the jet boundary layer of the reconnecting HCS, rather than on possible ion-scale structures embedded within the main body of the reconnection exhaust. The latter type of substructure was examined in a recent study by \citet{Phan2024}, who reported a reconnecting HCS containing multiple embedded reconnecting CSs inside the exhaust. The thicknesses of those subscale CSs ranged from $\sim$20 to $2000~d_i$, generally larger than the ion-scale CSs considered here. In our event, comparable ion-scale CSs may also be present within the HCS reconnection exhaust, as suggested by Figs.~\ref{Fig1} and \ref{Fig2} over the interval between $\mathrm{T}_5$ and $\mathrm{T}_6$. A dedicated analysis of such structures is beyond the scope of the present study and is deferred to future work.

\begin{acknowledgements}
This work was supported by the National Research Foundation of Korea (NRF) grant funded by the Korea government (MSIT) (RS-2024-00454886). We acknowledge the NASA Parker Solar Probe Mission and SWEAP team led by Justin Kasper for use of the data. The FIELDS experiment on the Parker Solar Probe spacecraft was designed and developed under NASA contract NNN06AA01C.
\end{acknowledgements}

\bibliographystyle{aa}
\bibliography{ref}

\end{document}